\documentstyle[twocolumn,aps,prb]{revtex}
\begin{document}
\input epsf.sty

\twocolumn[\hsize\textwidth\columnwidth\hsize\csname %
@twocolumnfalse\endcsname

\draft
\widetext

\title{Quantum Monte Carlo Study of Weakly Coupled Spin Ladders}
\author{Y. J. Kim}
\address{Division of Engineering and Applied Sciences, Harvard
University, Cambridge, Massachusetts 02138\\
and Center for Materials Science and Engineering,
Massachusetts Institute of Technology, Cambridge, Massachusetts 02139}
\author{R. J. Birgeneau, M. A. Kastner, and Y. S. Lee}
\address{Department of Physics and Center for Materials Science and 
Engineering, Massachusetts Institute of Technology, Cambridge, 
Massachusetts 02139} 
\author{Y. Endoh}
\address{Department of Physics, Tohoku University, Sendai 980-8578, Japan}
\author{G. Shirane}
\address{Department of Physics, Brookhaven National Laboratory, Upton, 
New York 11973} 
\author{K. Yamada}
\address{Department of Physics, Kyoto University, Gokasho, Uji 610-0011, 
Japan}

\date{\today}
\maketitle

\begin{abstract}

We report a quantum Monte Carlo study of the thermodynamic properties of
arrays of spin ladders with various widths ($n$), coupled via a weak
inter-ladder exchange coupling $\alpha J$, where $J$ is the intra-ladder
coupling both along and between the chains. This coupled ladder system
serves as a simplified model for the magnetism of presumed ordered spin
and charge stripes in the two-dimensional CuO$_2$ planes of hole-doped
copper oxides. Our results for $n=3$ with weak inter-ladder coupling
$\alpha=0.05$, estimated from the $t-t'-t''-J$ model, show good agreement
with the ordering temperature of the recently observed spin density wave
condensation in La$_2$CuO$_{4+y}$. We show that there exists a quantum
critical point at $\alpha_c \simeq 0.07$ for $n=4$, and determine the
phase diagram. Our data at this quantum critical point agree
quantitatively with the universal scaling predicted by the quantum
nonlinear $\sigma$ model. We also report results on random mixtures of
$n=2$ and $n=3$ ladders, which correspond to the doping region near but
above 1/8. Our study on the magnetic static structure factor reveals a
saturation of the incommensurability of the spin correlations around
1/8, while the incommensurability of the charge stripes grows linearly
with hole concentration. The implications of this result for the
interpretation of neutron scattering experiments on the dynamic spin
fluctuations in La$_{2-x}$Sr$_x$CuO$_4$ are discussed.

\end{abstract}

\phantom{.}
]

\narrowtext

\section{Introduction}
\label{sec:intro}

One of the most surprising early observations in the field of high
temperature superconductivity is that the onset of superconductivity as a
function of hole concentration coincides with a
commensurate-incommensurate transition in the low-energy dynamical spin
fluctuations.\cite{Endoh92} These incommensurate spin fluctuations have
been studied extensively over the past decade.\cite{Kastner98} We note
especially the recent detailed quantitative study by Yamada {\em et
al.}\cite{Yamada98} A variety of theoretical explanations of the spin
incommensurability have been offered, varying from nesting of the Fermi
surface in a nearly free electron model\cite{Bulut90,Si93,Tanamoto94} to
microscopic phase separation of the holes in a doped-Mott insulator model
for the copper oxides.\cite{Zaanen,Emery+Kivelson,Nayak97} Specifically,
in the latter model under certain circumstances the holes could organize
themselves into {\it stripes} with the charged stripes acting as antiphase
domain walls for the intervening antiferromagnetic regions.  The latter
correspond to spin ladders which have been recently studied quite
extensively in a different context.  \cite{Dagotto96,Greven96} None of the
experiments on the spin dynamics has been able to select unambiguously
between these various theoretical models. 

However, in the past two years, the experimental situation has changed
significantly.  First, rather dramatic elastic magnetic scattering effects
have been observed \cite{Tranquada} at low temperatures in samples of
La$_{1.6-x}$Nd$_{0.4}$Sr$_x$CuO$_4$ with $x$=0.12, 0.15, and 0.20. These
materials are all superconductors with onset $T_c$'s of $\simeq 4K$, $11K$
and $15K$, respectively.  In each case, Tranquada and coworkers observe
elastic incommensurate magnetic peaks with onset temperatures of $\simeq
50K$, $46K$ and $15K$, respectively.  Recent work \cite{Tranquada98b} has
shown that in the $x$=0.12 sample the correlation length reaches its
maximum value below $\sim 30K$. The incommensurabilities are essentially
identical to those of the corresponding dynamical fluctuations in samples
of La$_{2-x}$Sr$_x$CuO$_4$ with the same $x$.  In their study of the spin
dynamics in La$_{2-x}$Sr$_x$CuO$_4$, Yamada {\em et al.} observe that the
momentum space width of the low-energy spin fluctuations is a minimum for
hole concentrations near 1/8.  Further, at these hole concentrations the
spin fluctuations extend down to very low energies even in the
superconducting state.  This is in contrast to the situation in samples
with hole concentrations at and above the optimal doping level, $x \simeq
0.15$, where a well-defined spin gap is observed in the superconducting
state.  This has led Suzuki {\em et al.} \cite{Suzuki98} and more
recently, Kimura {\em et al.} \cite{Kimura98} to search for elastic
magnetic scattering effects in La$_{1.88}$Sr$_{0.12}$CuO$_4$. Remarkably,
they observe a spin density wave transition at $T_m \simeq 31K$ which
coincides to within the errors with the onset temperature for
superconductivity in this sample;  the incommensurability is
$\epsilon=0.12$ which equals the Sr-doping, $x$.  Kimura and coworkers
also observe magnetic order in a sample of La$_{1.90}$Sr$_{0.10}$CuO$_4$
($T_c=31K$) with incommensurability $\epsilon=0.105$ below $T_m \simeq
15K$. 

Most recently, Lee and coworkers \cite{Lee98} have searched for spin
density wave order in a sample of La$_{2}$CuO$_4$ doped with oxygen, that
is, La$_{2}$CuO$_{4+y}$, which has a superconducting $T_c$ (onset) of
$42K$ and is predominantly stage-4.  They indeed find a transition to
long-range incommensurate magnetic order at $T_m \simeq 42K$ with typical
Bardeen--Cooper--Schrieffer (BCS) mean field behavior of the order
parameter below $T_m$. Importantly, they find that both the spin ordering
direction and the three dimensional stacking arrangement coincide with
those in pure La$_{2}$CuO$_4$.  The magnetic structure can be modeled
quite well with $n=3$ stripes of the La$_{2}$CuO$_4$ in-plane structure
separated by $n=1$ non-magnetic antiphase domain walls;  this gives an
incommensurability $\epsilon=0.125$, close to the measured value.  The
measured long-range ordered moment is $\sim 0.15 \mu_B$ and the
inter-layer magnetic correlation length is $\sim 3$ CuO$_2$ layers. 

In our view, these observations, including especially this recent work on
La$_{2}$CuO$_{4+y}$, give credence to stripe models for the microscopic
magnetic structure.  The recent observation \cite{YBCO} of similar
incommensurabilities in the spin dynamics of underdoped
YBa$_2$Cu$_3$O$_{6.6}$ strengthens this argument. Of course, a great deal
of work remains to be done to explain all of the experimental
observations, especially the electronic properties in the normal and
superconducting states.  Kivelson {\it et. al.}\cite{Kivelson98} consider
the electronic degrees of freedom within the charge stripes and
characterize these phases as electronic liquid crystals. Although the
charge and spin degrees of freedom of the charged stripes themselves are
clearly essential to the physics, it is nevertheless important as a first
step to develop a deeper understanding of the simpler problem of the
magnetism of idealized insulating stripe arrays as a function of stripe
width or, equivalently, hole concentration.  Accordingly, we adopt a naive
model, in which for hole concentrations between $\sim 0.05$ and 0.15 there
is approximately one hole per two coppers \cite{Tranquada,Nayak96}
along site-centered charge
stripes running along the tetragonal Cu--O--Cu--O... axes with the holes
confined to single chains. We assume further that the
charge stripes are effectively non-magnetic and that the magnetic coupling
across the stripes is, to first order, determined by the
third-nearest-neighbor coupling within the CuO$_2$ plane.  This is
believed to be antiferromagnetic and to be about 5\% of the nearest
neighbor coupling. \cite{CKim98} We omit entirely any effects of the low
energy charge and spin excitations of the charged stripes themselves. In
this greatly simplified model, one may use standard quantum Monte Carlo
(QMC) techniques to determine the magnetic properties of the stripe
arrays. 

Our understanding of two-dimensional quantum magnetism has grown
enormously since the discovery of high temperature superconductivity.
Theoretical efforts to understand the underlying antiferromagnetism in the
parent compounds as well as experimental efforts to synthesize and study
new cuprate materials have both contributed significantly. In particular,
the two-dimensional quantum Heisenberg antiferromagnet (2DQHA) has been
studied extensively, since the pioneering work by Chakravarty, Halperin,
and Nelson.  \cite{Chakravarty} They have mapped the long-wavelength,
low-temperature behavior of the 2DQHA to a quantum nonlinear $\sigma$
model (QNL$\sigma$M) in (2+1) dimensions and have obtained a phase diagram
with three regimes; quantum disordered (QD), quantum critical (QC), and
renormalized classical (RC). They have argued that the 2DQHA on the square
lattice for $S \geq 1/2$ with nearest neighbor interactions is in the RC
regime, and consequently has a correlation length diverging exponentially
in $1/T$ as temperature is lowered to $T=0$, implying the existence of a
long-range ordered ground state. The temperature dependence of the
correlation length in the RC regime, which has been solved exactly to
three loop order by Hasenfratz and Niedermayer (HN), \cite{Hasenfratz91}
agrees quantitatively with the results of neutron scattering experiments
on both Sr$_2$CuO$_2$Cl$_2$ \cite{Greven95} and La$_2$CuO$_4$. 
\cite{Birgeneau95}

One-dimensional (1D) quantum magnetism has also drawn much interest in the
last decade, mainly due to a conjecture made by Haldane regarding the
different ground state properties between spin chains with integer and
half-integer spin quantum number. \cite{Affleck89b} In his seminal work in
1983, \cite{Haldane83a} Haldane mapped the 1D Heisenberg model onto the
QNL$\sigma$M and showed that the half-integer spin chain has an additional
topological term, which generates gapless low-energy excitations together
with algebraically decaying spin correlation functions. Integer spin
chains, however, can be described by the standard QNL$\sigma$M in (1+1)
dimensions, in which the ground state is disordered due to quantum
fluctuations and the excitation spectrum acquires a gap -- the so-called
{\it Haldane gap}. This conjecture has been subsequently confirmed both
numerically\cite{Nightingale86} and experimentally. \cite{Renard87} Spin
ladders have been studied mostly in the same context;\cite{Dagotto96} when
even numbers of $S=1/2$ chains are coupled to form a ladder (even-width),
they show the same universal behavior as integer spin chains, while
odd-width $S=1/2$ ladders behave essentially like a single $S=1/2$ chain
at low temperatures and long wavelengths.

\begin{figure} 
\centerline{\epsfxsize=3.1in\epsfbox 
{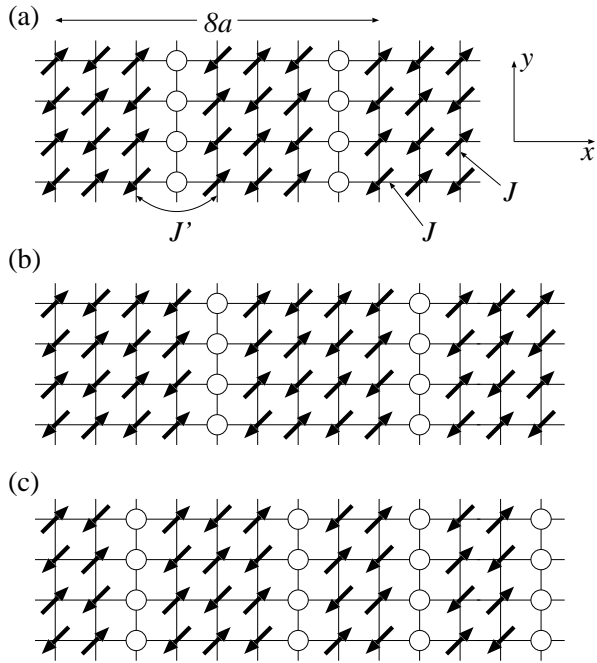}}
\vspace{0.1in} 
\caption{Schematic diagram of the model. (a) Weakly coupled
three leg ladder array with a periodicity of $8a$. $J'$ is the
inter-ladder coupling $\alpha J$. (b) Weakly coupled four leg ladder
array. (c) Random mixture of two and three leg ladders.} 
\label{fig1}
\end{figure}

In this paper we report a detailed QMC study of ladder arrays as a
function of the width of the spin ladders ($n$), the strength of the
coupling across the non-magnetic line ($J'=\alpha J$), and temperature
($T$).  We have carried out such calculations for $n=4$ ladder arrays,
$n=3$ ladder arrays, and for random mixtures of $n=3$ and $n=2$ ladders. 
These various arrays are illustrated in Fig.\ \ref{fig1}. In the limit of
$\alpha \rightarrow 0$, the model corresponds to a set of isolated
ladders, and in the limit $\alpha \rightarrow 1$ we recover the isotropic
$S=1/2$ square lattice QHA.  One can easily see that a quantum critical
point as a function of $\alpha$ should exist for non-zero $\alpha$ for
coupled even-width ladders. Indeed, the recent study by Tworzydlo {\em et
al.} \cite{Tworzydlo99} shows that $\alpha_c \simeq 0.3$ for coupled $n=2$
ladders. We show that arrays of $n=4$ ladders exhibit a quantum critical
point around $\alpha \simeq 0.07$. We discuss our QMC results in the
context of available experimental information;  we also suggest future
measurements which should determine whether or not these calculations are
in fact relevant to the real monolayer CuO$_2$ superconductors. 

The format of this paper is as follows:  In Sec.\ \ref{sec:QMC} we discuss
our QMC techniques.  Section\ \ref{sec:3leg} contains our results for
weakly coupled three-leg ladders ($n=3$).  Results for weakly coupled
four-leg ladders ($n=4$) including especially the quantum critical
behavior are given in Sec.\ \ref{sec:4leg}. In Sec.\ \ref{sec:mix}, we
discuss the results for random mixtures of two-leg and three-leg ladders.
We have also studied strongly coupled $n=3$ ladders and present the
results in Sec.\ \ref{sec:strong}. Finally, a discussion, summary, and
conclusions are given in Sec.\ \ref{sec:discussion}. 

\section{Quantum Monte Carlo}
\label{sec:QMC}

We have carried out quantum Monte Carlo simulations on large lattices
utilizing the loop cluster algorithm. \cite{Evertz98} The same algorithm
previously employed to study spin ladders \cite{Greven96} and chains
\cite{Kim98} is used with minor modifications. We obtain the temperature
dependence of the uniform susceptibility, $\chi_{\rm u}(T)$, the spin-spin
correlation length in the $x$-direction, $\xi_x(T)$, and the
$y$-direction, $\xi_y(T)$, the staggered susceptibility, $\chi_{\rm
s}(T)$, and the static structure factor at the antiferromagnetic wave
vector ${\bf Q}\equiv(\pi,\pi)$, $S_Q(T)$. The Hamiltonian for our model
is based on the square lattice with a coupling $J$, but every $n$-th bond
in the $x$-direction is replaced by a bond with exchange coupling $\alpha
J$: 
\begin{eqnarray} 
{\cal H} &=& J \left[ \sum_{i,j} {\bf S}_{i,j} \cdot
{\bf S}_{i,j+1} + \sum_{i \neq ln, j} {\bf S}_{i,j} \cdot {\bf S}_{i+1,j}
\right. \nonumber\\ 
&+& \left. \alpha \sum_{i=ln,j} {\bf S}_{i,j} \cdot
{\bf S}_{i+1,j} \right], 
\label{hamil} 
\end{eqnarray} 
where $i$ and $j$
run over the $x$ and $y$ directions, respectively. $n$ is the width of the
ladder and $l$ counts the ladders. We use units in which $\hbar=k_B=g
\mu_B=1$. We also set the lattice constant $a=1$. Periodic boundary
conditions are used in both directions. 

One should note that, although this Hamiltonian captures the essential
magnetic interactions of coupled ladders, if Eq.\ (\ref{hamil}) is to
describe the stripe arrays shown in Fig.\ \ref{fig1}, the lattice constant
in the $x$-direction is $\frac{n}{n+1}$ instead of one. This affects only
two quantities in our computation; $\xi_x(T)$ and $S_Q(T)$. The correction
for $\xi_x(T)$ is trivial, since $\xi_x(T)$ is measured in units of the
lattice constant.  Unless noted otherwise, this $\frac{n}{n+1}$ correction
is incorporated in the $\xi_x$ data presented in this paper.  $S_Q(T)$,
however, cannot be converted in a simple manner for the stripe model.
Therefore, in the following sections, $S_Q(T)$ should be understood as the
structure factor at $(\pi,\pi)$ for the model Hamiltonian, Eq.\
(\ref{hamil}), rather than for the stripe model as shown in Fig.\
\ref{fig1}. The incommensurate structure factor is discussed only in Sec.\
\ref{sec:mix}, where it is separately computed as a function of $q_x$,
$S(q_x,\pi)$. 

In order to deduce the spin-spin correlation lengths, $\xi_x(T)$ and
$\xi_y(T)$, the instantaneous spin-spin correlation function, $C(x,y)$, is
computed and fitted to the asymptotic form 
\begin{mathletters}
\begin{eqnarray} 
C(x,0) \sim x^{-\lambda} e^{-x/\xi_x} 
+ (L_x-x)^{-\lambda} e^{-(L_x-x)/\xi_x}\label{OZ-x}\\ 
C(0,y) \sim y^{-\lambda} e^{-y/\xi_y} 
+ (L_y-y)^{-\lambda} e^{-(L_y-y)/\xi_y},
\label{OZ-y} 
\end{eqnarray} 
\end{mathletters}
which is a symmetrized 2D Ornstein-Zernike form
($\lambda=0.5$). Only data with $x \agt 3\xi_x$ and $y \agt 3\xi_y$ are
included in the fits to ensure that the asymptotic behavior is probed. 
Technically, due to the non-uniformity of the exchange couplings in the
$x$-direction, $C(x,0)$ can be better described by the function $\sim
x^{-1/2}e^{-x/\xi_x} \sin(x/n)$ rather than Eq.\ (\ref{OZ-x}); however, the
values for $\xi_x$ obtained with both methods agree within the error bars. 

The lengths and Trotter numbers of the simulated lattices are chosen so as
to minimize any finite-size and lattice-spacing effects.  The linear sizes
of the lattice, $L_x$ and $L_y$, are kept at least 10 times larger than
the respective correlation lengths. Spin states are updated about $2
\times 10^{4}$ times to reach equilibrium and then measured $\sim 5 \times
10^{4}$ times. For the random mixtures of $n=2$ and $n=3$ ladders, we
generate each configuration in the following way: Beginning from the first
column ($i=1$), we assign a spin or hole for each column; the number of
spin columns between the hole columns represents the width of the spin stripe.
In order to make sure that the width is either two or three, we put
restrictions such that a spin column must be assigned after each hole
column, while a hole column must be assigned after three consecutive spin
columns. The width of each ladder is determined by generating a random
number $r$ ($0<r<1$) and comparing $r$ with $p$, where $p$ is the ratio of
the three leg ladder in the mixture. After two spin columns are assigned,
the spin column is assigned only if $r<p$, otherwise the hole column is
assigned and the width becomes two. For example, for an equal mixture of
two and three leg ladders, $p=0.5$. For each $T$ and $p$, typically 10 to
20 different configurations are generated and averaged over. Averaging
over more configurations does not alter our results.

\section{Weakly coupled Three Leg Ladders}
\label{sec:3leg}

\begin{figure}
\centerline{\epsfxsize=3.2in\epsfbox
{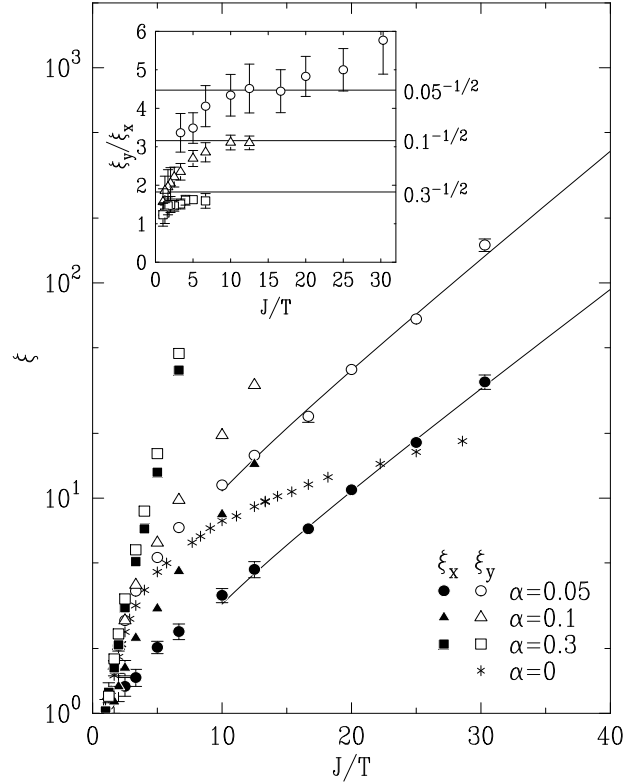}} 
\vspace{0.1in}
\caption{Correlation lengths for arrays of 
$n=3$ ladders with different inter-ladder
coupling $\alpha$ plotted as a function of inverse temperature.
Correlation lengths in the $x$-direction are shown as solid symbols, while
those in the $y$-direction are shown as empty symbols. For $\alpha=0.05$,
the fitting results from Eq.\ (\ref{CastroNeto}) are also plotted. Inset:
The ratio of $\xi_y /\xi_x$. Each solid line corresponds to the mean field
value, $\sqrt{\alpha}$. The ratio is given without the $n/(n+1)$
correction discussed in Sec.\ \ref{sec:QMC}.} 
\label{fig2} 
\end{figure}

We present our QMC data for the $n=3$ case, Fig.\ \ref{fig1}(a), in this
section. From detailed studies of isolated ladders, \cite{Greven96} it is
known that at low temperatures and long length scales three-leg ladders
exhibit the same behavior as a single chain ($S=1/2$).  It is well known
that any non-zero inter-ladder coupling eventually enhances the
correlations across the non-magnetic stripe and drives the system towards
the 2DQHA-QNL$\sigma$M RC fixed point. Thus, we expect to see
qualitatively different behavior for an array of weakly coupled $n=3$
ladders from that of isolated ladders. The correlation lengths for arrays
of $n=3$ ladders are shown in Fig.\ \ref{fig2} as a function of inverse
temperature for various inter-ladder couplings ($\alpha$). Note that
$\alpha=0$ corresponds to an isolated three leg ladder; these data are
taken from Ref.\ \onlinecite{Greven96}.  Not surprisingly, one can see
clearly $\sim \exp(1/T)$ behaviors except for $\alpha=0$, as discussed
above. In the inset of Fig.\ \ref{fig2}, we also show the ratio $\xi_y /
\xi_x$. At low temperatures, $\xi_y/\xi_x$ approaches the mean field
prediction, $\sqrt{\alpha}$, shown here as a solid line for each $\alpha$. 

Our correlation length data are fitted to the crossover form given by
Castro Neto and Hone, \cite{CastroNeto96} which interpolates between the
HN result at low temperatures and $\xi \sim T^{-1}$ at high temperatures:
\begin{equation} 
\xi= A \exp({2 \pi \rho_s \over T}){1 \over {1+{1 \over
2}{T \over 2 \pi \rho_s}}}, 
\label{CastroNeto} 
\end{equation} 
where $\rho_s$ is the spin stiffness.  Our fitting
results are presented in Table\ \ref{table1}. We note first that the
values deduced for $2 \pi \rho_s$ separately from $\xi_y$ and $\xi_x$ are
the same within the error bars, which means that there is only one
temperature scale for the low-temperature behavior of this model, despite
the fact that the correlation lengths themselves are highly anisotropic. 

\begin{table}
\caption{Fitting parameters for n=3 data to Eq.\ (\ref{CastroNeto})
without the $n/(n+1)$ correction. 
\label{table1}}
\begin{tabular}{ccccc}
        &\multicolumn{2}{c}{$2\pi \rho_s/J$}&\multicolumn{2}{c}{$A$}\\
$\alpha$ & $\xi_x$ & $\xi_y$ & $\xi_x$ & $\xi_y$\\
\tableline
0.05    & 0.10(1) & 0.11(1) & 0.26(2) & 1.14(5)\\
0.1    & 0.17(2) & 0.19(1) & 0.52(5) & 1.4(1)\\
0.3    & 0.55(2) & 0.58(1) & 0.83(5) & 1.22(5)\\
1.0\tablenote{Ref.\ \onlinecite{Beard98}}     & \multicolumn{2}{c}{1.1310(3)} 
& \multicolumn{2}{c}{0.4978(8)}\\
\end{tabular}
\end{table}

\begin{figure}
\centerline{\epsfxsize=3.2in\epsfbox
{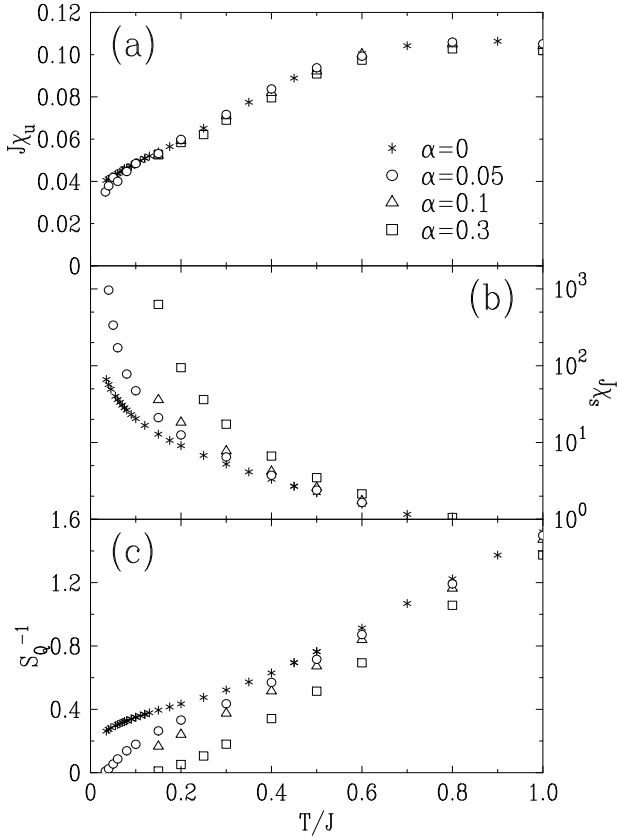}}
\vspace{0.1in}
\caption{Uniform susceptibility, staggered susceptibility, and inverse
static structure factor peak intensity for arrays of $n=3$ ladders with different
inter-ladder couplings $\alpha$. }
\label{fig3}
\end{figure}

Since the 2D Heisenberg model cannot have long-range order at any non-zero
temperature, either a small anisotropy in the spin Hamiltonian or a
non-zero 3D coupling is necessary to explain the N\'eel transition in real
materials. In pure La$_2$CuO$_4$, one can estimate the N\'eel transition
temperature reliably as the temperature at which the correlation length
squared becomes of order of the inverse of the effective anisotropy,
\cite{Keimer92b} which, in the case of La$_2$CuO$_4$, is about
$\gamma_{eff} \approx 10^{-4}$. We can apply the same heuristic formula to
determine the spin density wave ordering temperature, $T_m$, for our
presumed stripe array model for doped La$_2$CuO$_4$.  For example,
La$_2$CuO$_{4+y}$ has the same spin ordering direction and stacking scheme
as in pure La$_2$CuO$_4$ except that the inter-layer order extends over
only about three planes; \cite{Lee98} thus, we take the same effective
anisotropy and find the temperature satisfying $\xi_x(T_m) \xi_y(T_m)
\approx \gamma_{eff}^{-1}$. In our simplified model the inter-ladder
coupling is determined by the third-nearest-neighbor exchange coupling in
the CuO$_2$ plane;  from current values for $t,t',t''$ and $J$ in the
extended $t-J$ model \cite{CKim98} we estimate $\alpha \approx (t''/t)^2
\approx 0.05$. By extrapolating our correlation length data to lower
temperature for $\alpha=0.05$, we obtain $T_m \approx 0.029J$, or $T_m
\approx 44 K$ for $J=1500K$, which agrees remarkably well with the
experimental result. This precise agreement is a coincidence since $\alpha
= 0.05$ is only a crude estimate and we have assumed perfectly ordered
charge stripes. We note that $\alpha=0.1$ gives $T_m \approx 75 K$. We
should also note that these estimates of $T_m$'s are not very sensitive to
the explicit value of $\gamma_{eff}$. Even if $\gamma_{eff}$ changes by a
factor of two, $T_m$ remains within 10\% of the value obtained here.
However, $T_m$ depends sensitively on the choice of the inter-ladder
exchange coupling $\alpha J$. The low values for $T_m$ found
experimentally clearly constrain the inter-ladder coupling to rather small
values (or, as discussed in Sec.\ \ref{sec:strong}, very large values), at
least within the simplified model considered here.

In Fig.\ \ref{fig3}, Monte Carlo results for $\chi_u$, $\chi_s$, and the
inverse of $S_Q$ are plotted as a function of $T$, for different
$\alpha$'s.  The reason $S_Q^{-1}$ is plotted is to emphasize the effect
of the inter-ladder coupling.  One of the most significant results of
Ref.\ \onlinecite{Kim98} is that the divergence of $S_Q$ is only
logarithmic in 1D; thus $S_Q^{-1}$ does not extrapolate to zero at zero
temperature. As shown in Fig.\ \ref{fig3}(c), coupled $n=3$ ladders behave
completely differently from the isolated $n=3$ ladder; namely, the $S_Q$
of coupled ladders exhibits a clear crossover from the weak divergence in
$T^{-1}$ of an isolated ladder to the strong divergence in $T^{-1}$ of a
2D spin system. 

\begin{figure}
\centerline{\epsfxsize=3.2in\epsfbox
{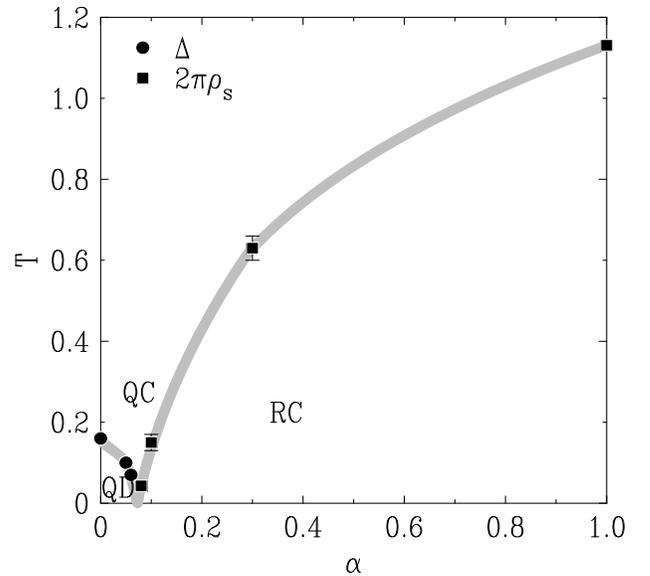}}
\vspace{0.1in}
\caption{Phase diagram for weakly coupled four leg ladders. The quantum 
critical point corresponds to $\alpha=0.07(1)$, as shown in Fig.\ 
\ref{fig7}. The shaded line represents a crossover region rather than a 
true phase boundary. }
\label{fig4}
\end{figure}

\section{Weakly coupled Four Leg Ladders}
\label{sec:4leg}

Results for $n=4$, corresponding to Fig.\ \ref{fig1}(b), are presented in
this section.  Unlike the $n=3$ case, an isolated four-leg ladder has a
Haldane gap ($\Delta \approx 0.16J$). Therefore, if $\alpha J \ll \Delta$,
the ground state remains disordered, and there must then be a quantum
phase transition as a function of $\alpha$ at zero temperature at a
critical value $\alpha_c$ to an ordered ground state.  In QNL$\sigma$M
language, there is a quantum critical point dividing the QD ground state
and the RC ground state at $\alpha_c$. We show the phase diagram of arrays
of $n=4$ ladders in Fig.\ \ref{fig4}. In this figure, we take $2 \pi
\rho_s$ as the temperature scale where crossover from the QC to the RC
regime occurs, and $\Delta$ as the crossover temperature from the QC to
the QD regime. $2 \pi \rho_s$ is determined in the manner explained in
Sec.\ \ref{sec:3leg}; $\Delta$ can be obtained by fitting to
$\chi_u(T)\sim \exp(-\Delta/T)$ at low temperatures ($T \alt \Delta$). One
should note that the thick shaded lines in Fig.\ \ref{fig4} represent
crossover lines rather than true phase boundaries. The phase diagram shown
here is basically that of the QNL$\sigma$M, \cite{Chakravarty}
with $\alpha$ playing the role
of the coupling constant $g$. As shown in Fig.\ \ref{fig4}, a quantum
critical point occurs at $\alpha_c =0.07(1)$. 

\begin{figure} 
\centerline{\epsfxsize=3.2in\epsfbox
{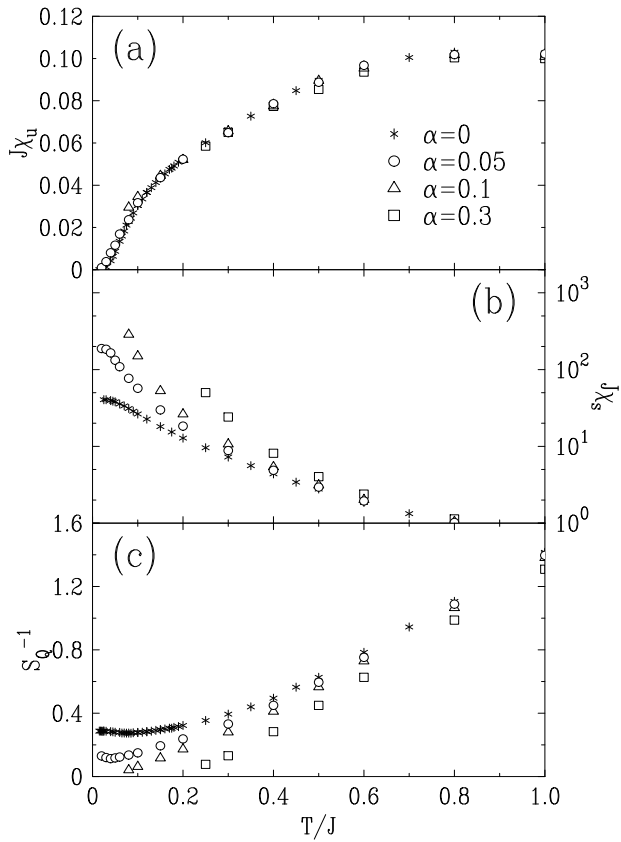}}
\vspace{0.1in}
\caption{Uniform susceptibility, staggered susceptibility, and inverse 
static structure factor peak intensity for arrays of 
$n=4$ ladders with different 
inter-ladder couplings $\alpha$. }
\label{fig5}
\end{figure}

\begin{figure} 
\centerline{\epsfxsize=3.2in\epsfbox
{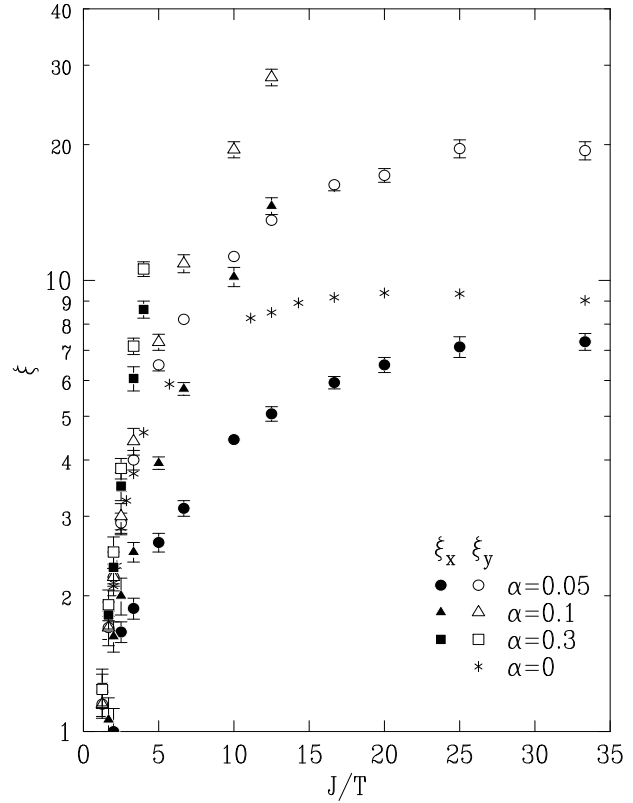}}
\vspace{0.1in}
\caption{Correlation lengths for arrays of 
$n=4$ ladders with different inter-ladder
couplings $\alpha$ plotted as a function of inverse temperature.
Correlation lengths in the $x$-direction are shown as solid symbols, while
those in the $y$-direction are shown as empty symbols. }
\label{fig6}
\end{figure}

Monte Carlo results for $\chi_u$, $\chi_s$, and $S_Q^{-1}$ are plotted as
a function of $T$, for different $\alpha$'s, in Fig.\ \ref{fig5}. The
correlation length obtained from the simulation is plotted in Fig.\
\ref{fig6} as a function of $T^{-1}$. Figures\ \ref{fig5} and \ref{fig6}
are plotted in such a way as to contrast their behaviors with those of
arrays of $n=3$ ladders.  The difference between the behaviors in the QD
and the RC regime is evident in these figures;  namely, $\xi$, $\chi_s$,
and $S_Q$ all diverge exponentially at low temperatures in the RC regime
(e.g., $\alpha=0.3$), while they all saturate at finite values as $T
\rightarrow 0$ in the QD regime (e.g., $\alpha=0.05$). 

\begin{figure}
\centerline{\epsfxsize=3.2in\epsfbox
{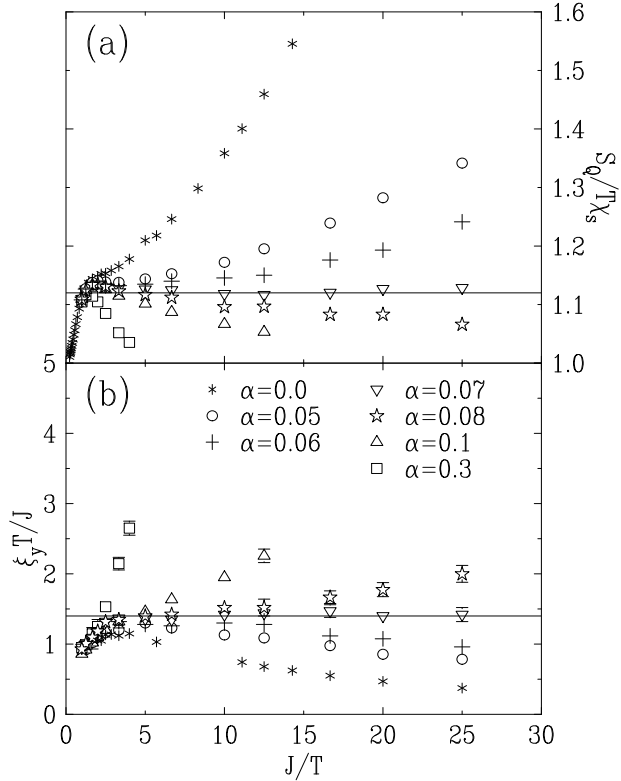}}
\vspace{0.1in}
\caption{ (a) The ratio $S_Q /T \chi_s$ as a function of inverse
temperature as described in the text. (b) The correlation length in the
$y$-direction of arrays of 
$n=4$ ladders multiplied by temperature is plotted to
contrast the different low-temperature behaviors in the QD and RC phases. 
The solid lines in both (a) and (b) are the predictions for quantum
critical scaling. } 
\label{fig7}
\end{figure}

To illustrate the quantum critical behavior more dramatically, we plot in
Fig.\ \ref{fig7}(a), the dimensionless ratio $S_Q/T \chi_s$. At very high
temperatures, this ratio shows classical behavior, $S_Q/T \chi_s =1$,
while the same behavior shows up again at very low temperatures for
$\alpha > \alpha_c$, that is, in the RC regime. This is as expected, since
RC behavior is closely similar to classical behavior, if the spin wave
velocity and spin stiffness renormalizations are accounted for. In the QD
regime, that is, $\alpha < \alpha_c$, $S_Q$ and $\chi_s$ are constant;
therefore $S_Q/T \chi_s$ should be linear in $T^{-1}$. According to the
quantum critical scaling predicted for the QNL$\sigma$M,
\cite{Chubukov,Sokol94} this ratio should show universal behavior in the
QC regime with the specific value: $S_Q/T \chi_s = 1.10(2)$. As may be
seen in Fig.\ \ref{fig7}(b), the $\alpha=0.07$ data for $S_Q/T \chi_s$
indeed are constant $\simeq 1.12$ at low temperatures, in quantitative
agreement with the QC theory. This strongly supports the above claim that
$\alpha_c=0.07$ for the $n=4$ stripe array system is a quantum critical
point. Another quantity plotted in Fig.\ \ref{fig7}(b) is $\xi_y T$.
Although $\xi_x T$ is not shown here, it shows essentially the same
temperature dependence. Again at very high temperatures, this quantity
shows classical behavior, $\xi_y T=J$, while the low-temperature
behavior can distinguish the QC regime from the RC and QD regimes. In the
QC regime, the QNL$\sigma$M \cite{Chubukov} predicts $\xi T = c/1.04$,
where $c$ is the spin wave velocity.  As may be seen in Fig.\
\ref{fig7}(b), $\xi_y T$ is indeed a constant at low temperatures (high
$1/T$) for $\alpha=0.07$; the explicit value $\xi_y T \simeq 1.4 J$
corresponds to $c_y \simeq 1.46J$, close to, but somewhat less than, the
value $c=1.657J$ for the square lattice. \cite{Beard98}

\section{Random Mixture of Weakly coupled two and three leg ladders}
\label{sec:mix}

\begin{figure}
\centerline{\epsfxsize=3.2in\epsfbox
{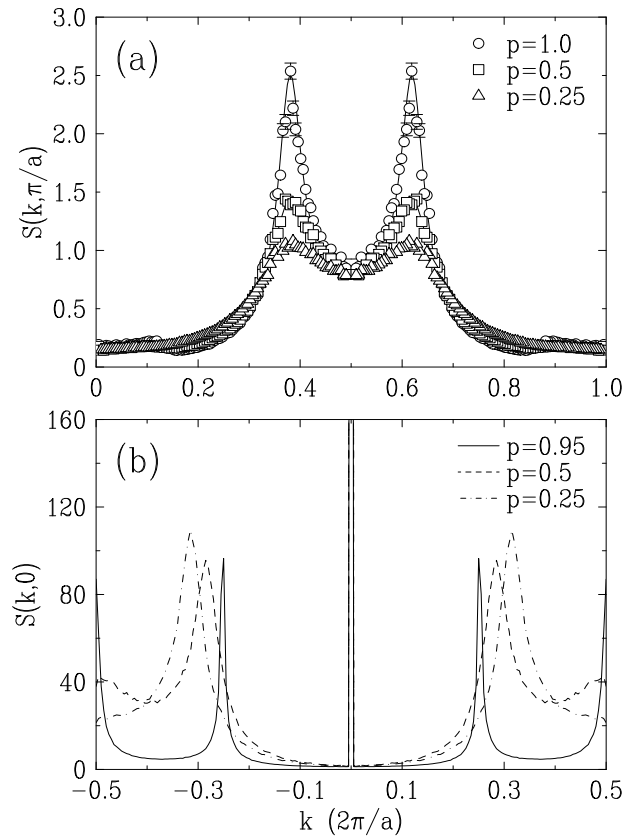}} 
\vspace{0.1in}
\caption{(a) Static Structure factor near the antiferromagnetic wave
vector for the $n=2$ and $n=3$ mixed spin ladders. Representative data at
$T=0.05$ for $p=0.5$ and $p=0.25$, and at $T=0.08$ for $p=1.0$ are shown;
$\alpha=0.05$ for all data.  The solid lines are the results of fits to
two Lorentzians [Eq.\ (\ref{structure_factor})].  (b) Corresponding charge
structure factor near the nuclear zone center, calculated as described in
the text. We show $p=0.95$ data instead of $p=1.0$ for graphical purpose.}
\label{fig8}
\end{figure}

\begin{figure}
\centerline{\epsfxsize=3.2in\epsfbox
{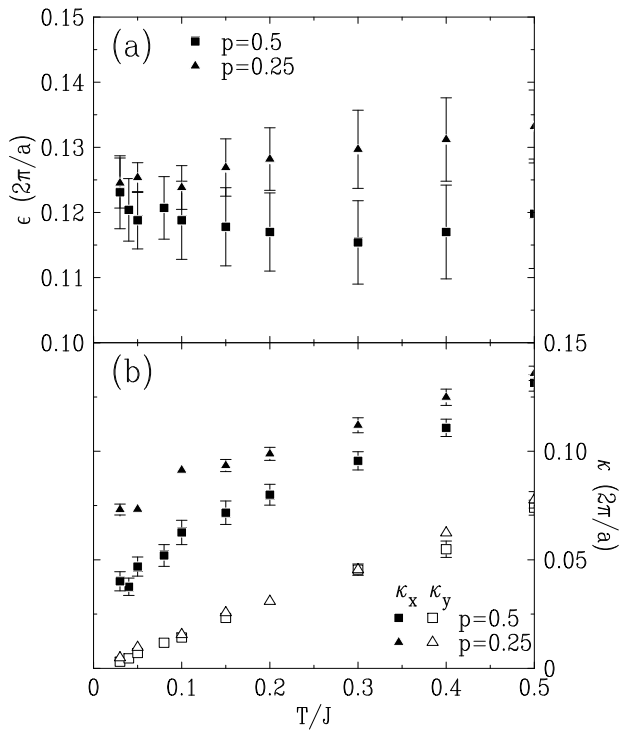}} 
\vspace{0.1in}
\caption{(a) The incommensurability obtained from the fitting static
structure factor to two Lorentzians together with a fixed background term
[Eq.\ (\ref{structure_factor})].  Even if there is a substantial fraction
of $n=2$ ladders, the incommensurability remains fixed near 1/8. (b) Inverse
correlation length as a function of $T$. $\kappa_x$ is obtained from the
fit. $\kappa_y$ is determined from the correlation function Eq.\
(\ref{OZ-y}) as described in Sec.\ \ref{sec:QMC}.} 
\label{fig9} 
\end{figure}

In Fig.\ \ref{fig8}(a), we show representative static structure factor
data at low temperatures for random mixtures of $n=2$ and $n=3$ weakly
coupled ladders, obtained as described in Sec.\ \ref{sec:QMC}.  Here, $p$
is the fraction of $n=3$ ladders in the mixture; therefore, $p=1.0$
corresponds to the pure $n=3$ case. We choose $\alpha=0.05$ for all of the
simulations discussed in this section. We show data at low temperatures
($T \sim 0.05J$), since the magnetic structure factor develops well
defined peaks only at these low temperatures. At high temperatures, due to
the short correlation lengths, the structure factor exhibits only broad
peaks, making it difficult to extract meaningful values for the
incommensurability.  Nevertheless, we can fit the data with two identical
Lorentzians, split symmetrically about the antiferromagnetic wave vector
${\bf Q} \equiv (\pi,\pi)$, together with a broad temperature-independent
background term centered around $(\pi,\pi)$: 
\begin{eqnarray} 
S(q_x,\pi)
&=& \frac{S_Q}{4}\left(\frac{1}{1+q_{+}^2/\kappa_x^2}
+\frac{1}{1+q_{-}^2/\kappa_x^2} \right) \nonumber\\ 
&+& \frac{B}{1+(q_x-\pi)^2/\kappa_b^2}. 
\label{structure_factor}
\end{eqnarray} 
Here $q_\pm=q_x-(\pi \pm \epsilon)$ and the last term is a
temperature-independent background. The width of this background term
corresponds to $\sim 1$ lattice constant.  There are only two temperature
dependent fitting parameters, $\epsilon$ and $\kappa_x$, since the peak
intensity $S_Q/4$ is calculated separately in our QMC study.  
The peak position corresponds to the
incommensurability $\epsilon$, and the peak width corresponds to the
inverse correlation length $\kappa_x$. These data are presented in Fig.\
\ref{fig9}. In Fig.\ \ref{fig9}(b), we also show $\kappa_y=1/\xi_y$;
$\xi_y$ is computed directly in the way explained in Sec.\ \ref{sec:QMC}. 

The most prominent feature in Fig.\ \ref{fig8}(a) is that the
incommensurability $\epsilon$ of the magnetic structure factor does not
change significantly as $p$ is varied. This is verified quantitatively in
Fig.\ \ref{fig9}(a), where it may be seen that the values for $\epsilon$
for both $p=0.5$ and $p=0.25$ cluster around 1/8, the exact value for
$p=1.0$.  Note that the high-temperature data have large error bars, since
the peaks are very broad. This behavior of the magnetic structure factor
may be contrasted with that of the charge structure factor. We show in
Fig.\ \ref{fig8}(b) the charge structure factor for the charge stripes in
the random mixtures. The calculation is made for the simplest charge
distribution: we assume that the scattering is unity from the charge
stripes (anti-phase domain walls) and zero from the spin ladders. Although
greatly oversimplified, as an illustration this calculation nevertheless
provides useful intuitive guidance. The most important qualitative result
is that the incommensurability of the charge stripes is not equal to twice
the magnetic incommensurability in the random mixtures. For example, the
$p=0.25$ case has a charge incommensurability of 0.31, which is quite
different from $2\epsilon=0.25(2)$, where $\epsilon$ is the magnetic
incommensurability. 

\begin{figure}
\centerline{\epsfxsize=3.2in\epsfbox
{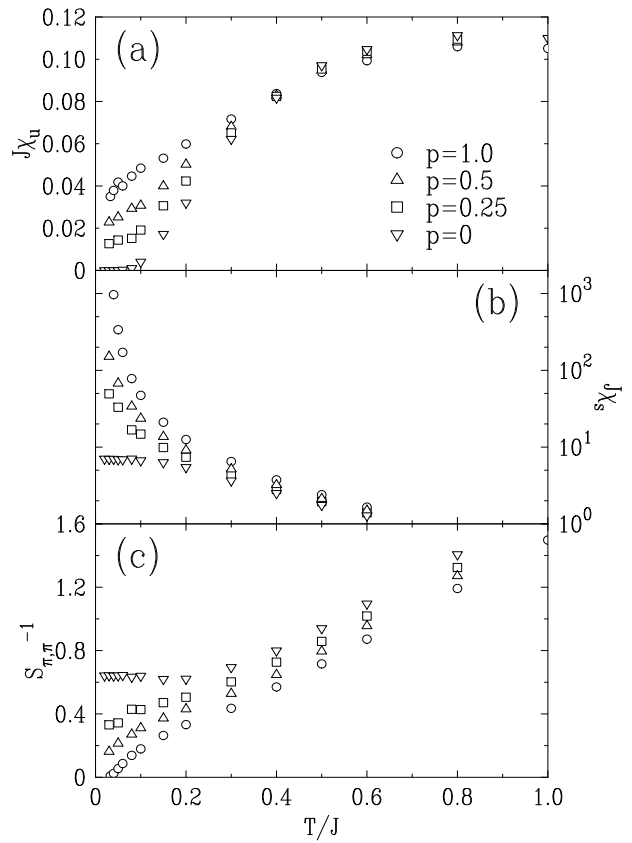}}
\vspace{0.1in}
\caption{Uniform susceptibility, staggered susceptibility, and inverse
static structure factor peak intensity for arrays of $n=2$ and $n=3$ mixed
ladders with different $p$. }
\label{fig10}   
\end{figure}

This saturation of $\epsilon$ of the magnetic structure factor around 1/8
can be understood in the following way: At low temperatures ($T \ll
\Delta$, $\Delta \approx 0.41J$) the spins on the two-leg ladder will form
a spin singlet ground state with a large spin gap in the excitation
spectrum. These spin singlets have effective spin zero, and therefore do
not contribute in first order to the magnetic structure. Therefore, the
magnetic structure factor originates predominantly from the $n=3$
components of the mixture.  In their systematic study of the {\it dynamic}
incommensurate spin fluctuations in La$_{2-x}$Sr$_x$CuO$_4$, Yamada and
coworkers\cite{Yamada98} showed that the incommensurability is linear in
hole concentration $x$ with $\epsilon \simeq x$ for $0.06 \alt x \alt
0.12$ and saturates around 1/8 on further doping. In our simplified model
of stripes, the range $0.12 \alt x \alt 0.16$ corresponds to mixtures of
$n=2$ and $n=3$ ladders. Although we have calculated the {\it static}
structure factor, we believe that the saturation in the dynamic
fluctuation incommensurability in La$_{2-x}$Sr$_x$CuO$_4$ may have the
same origin. We also have computed $\chi_u$, $\chi_s$, and $S_Q^{-1}$,
which are plotted in Fig.\ \ref{fig10} as functions of $T$, for different
$p$'s, all with $\alpha=0.05$. Note the differing low-temperature
behaviors of the $p=0$ arrays from those with $p \neq 0$; it is evident
that the $n=3$ physics indeed dominates in the magnetism of the random
mixture. 

Finally, we should note that in our model, in order to explain the fact
that the spin incommensurability saturates near 1/8 even for very high
dopings, we must hypothesize that for doping above the optimal value $x
\simeq 0.15$, the charge per stripe increases progressively with
increasing $x$ above $\sim 0.15$. That is, we assume that as the doping
increases beyond $x=0.15$ charge stripes cannot become closer due to the
Coulomb repulsion; instead, additional charges go into already existing
charge stripes. 

\section{Strongly coupled three leg ladders}
\label{sec:strong}

The previous calculations have all been done in the weak inter-ladder
coupling limit;  indeed, in order to obtain reasonable magnetic ordering
temperatures in our model, the reduced inter-ladder coupling must be of
order $\alpha \sim 0.05$. However, as pointed out to us by Kivelson,
\cite{Kivelson-private} a similar situation also should obtain in the
limit of large $\alpha$. In this case, physically one would imagine the
carriers along the charge stripes mediating a large effective
antiferromagnetic exchange between the bounding spin chains; these
neighboring chains would then form an array of two-leg ladders coupled to
single chains [Fig.\ \ref{fig1}(a)].

\begin{figure}
\centerline{\epsfxsize=3.2in\epsfbox
{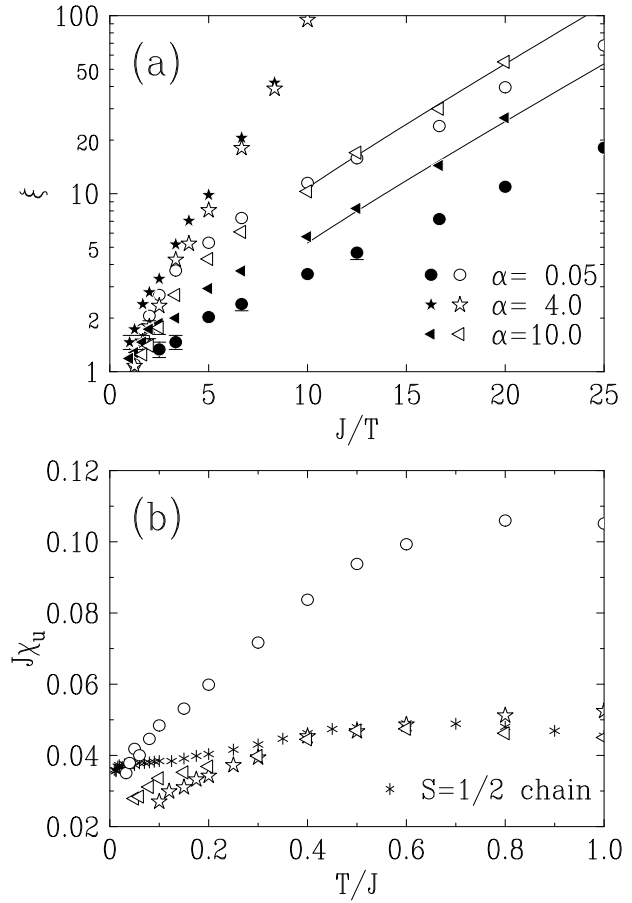}}
\vspace{0.1in}
\caption{(a) Correlation lengths for arrays of $n=3$ ladders with strong
inter-ladder coupling ($\alpha > 1$) plotted as a function of inverse
temperature. The correlation lengths in the $x$-direction are shown as
solid symbols, while those in the $y$-direction are shown as empty
symbols. (b) Uniform susceptibility per spin as a function of $T$.}
\label{fig11}  
\end{figure}

Accordingly, we have carried out a limited number of calculations on
arrays of $n=3$ ladders coupled with strong exchange couplings; that is,
in the $\alpha > 1$ limit.  Specifically, we have repeated our simulations
of Sec.\ \ref{sec:3leg} with $\alpha=4.0$ and $\alpha=10.0$. The
correlation length data so-obtained are shown in Fig.\ \ref{fig11}(a).  To
facilitate a comparison, we have also plotted the $\alpha=0.05$ data from
Fig.\ \ref{fig2}. At low temperatures, one can see the RC behavior
clearly, similar to that of the weak coupling data. The solid lines are
the results of fits to Eq.\ (\ref{CastroNeto}). It is evident, however,
that the uniform susceptibility data for strongly coupled $n=3$ ladders
shown in Fig.\ \ref{fig11}(b) are quite different from those of weakly
coupled ($\alpha=0.05$) $n=3$ ladders. In this limit, since $J'$ is the
primary coupling, as discussed above, two-leg ladders are effectively
formed across the charge stripes; therefore, we should consider this
system as a mixture of $S=1/2$ chains and $n=2$ ladders rather than as an
array of $n=3$ ladders. Specifically, $S=1/2$ chains and $n=2$ ladders
alternate in the $x$-direction. The spin gap of these $n=2$ ladders is
large (of order $0.5J'$) and the contribution from these to $\chi_u$ is
essentially zero. Thus one might surmise that $\chi_u$ would be entirely
due to the $S=1/2$ chains.  That this idea is correct may be seen from
Fig.\ \ref{fig11}(b), where we have plotted the uniform susceptibility 
per spin of the $S=1/2$ chain
from Ref.\ \onlinecite{Kim98} divided by three. Both the $\alpha=4.0$ and
the $\alpha=10.0$ data agree with the spin chain results for $T \agt
0.4J$. At low temperatures, the correlations between the chain and the
ladder become important and we observe behavior similar to that of
mixtures of ladders [Fig.\ \ref{fig10}(a)]. 

The correlation length for $\alpha=4.0$ grows as a function of $T^{-1}$
much faster than that of $\alpha=0.05$, such that the prediction for $T_m$
in this case is $T_m \approx 144 K$ for $J=1500K$, which is more than a
factor of three larger than the experimental result. In order to obtain
reasonable values for $T_m$ in the large $\alpha$ limit, one needs to have
$\alpha$ close to 10 as shown in Fig.\ \ref{fig11}. In such a case, for
$\alpha=10.0$, we obtain $T_m \approx 55 K$. Such large values for $J'
\approx 1.3$eV seem to us to be unlikely.  For this model ($J' \gg J$) to
be relevant one would need to invoke some other mechanism such as
extensive structural disorder or effects from charge and spin fluctuations
of the charged stripes themselves to reduce $T_m$ down to the observed
values of $30K$ to $40K$. 

\section{discussion}
\label{sec:discussion}

A few comments on the applicability of our model are in order. First and
foremost, our model is clearly an extremely simplified version of any
actual charge and spin ordering in the doped copper oxide planes. We
assume that (1) the charge stripes are perfectly ordered with the charges
confined to Cu--O--Cu--O chains and (2) the charge and spin degrees
of freedom along the stripes can be ignored. We know that dynamic
transverse fluctuations of the charge stripes could be very large,
especially for incommensurate charge stripe spacings. \cite{Kivelson98}
However, this effect may be less significant if the charge ordering occurs
at a much higher temperature than the spin ordering, so that the charge
fluctuations are significantly reduced near the spin ordering temperature.
Several neutron scattering experiments on La$_{2-x}$Sr$_x$NiO$_4$, and
La$_{1.6-x}$Nd$_{0.4}$Sr$_x$CuO$_4$ indeed report a charge ordering
temperature that is higher than the spin ordering temperature. 
\cite{Tranquada98a} The charge degrees of freedom in the metallic stripes,
which we have ignored, clearly are essential for the transport and
superconducting properties, \cite{Kivelson98} and we have nothing to
contribute on this aspect of the problem. Second, microscopic phase
separation of holes and spins can occur in other geometries; for example,
it is possible to have {\it grid-like} spin-rich regions separated by
hole-rich domain walls as well as diagonal stripes. In both cases, the
incommensurate spin density wave peaks would be rotated by 45$^\circ$ with
respect to those for the stripe model discussed here. 
In a recent experiment on an
$x=0.05$ sample, \cite{Wakimoto98} which notably is an insulator rather
than a superconductor at low temperatures, 45$^\circ$ rotated magnetic
peaks are indeed observed; in that case it is believed that this rotation
is due to the establishment of diagonal stripes.

Despite the success of the current model in explaining a number of
experimental observations for incommensurabilities near $1/8$, many points
remain to be understood.  As is obvious in Figs.\ \ref{fig2}, \ref{fig5},
and \ref{fig9}, $\xi_x$ and $\xi_y$ are very anisotropic, which would, in
turn, predict anisotropic widths in the quasielastic and dynamic neutron
scattering measurements. Currently available experimental data are not
complete enough to test this prediction. 
In recent work, Tranquada {\it et
al.}\cite{Tranquada98b} calculated the {\it elastic} structure factor of
$n=4$ and $n=3$ mixtures for sinusoidally varying charge and spin density
waves; they hypothesized that the stripe disorder is caused by pinning by
the random Sr dopants. They also argued that these defects would explain
the absence of any peak width anisotropy in their data. 

An obvious major deficiency of the current model is that it does not
naturally explain the special role which hole dopings $x$ and
concomitantly, incommensurabilities $\epsilon$ near $1/8$ play in the
La$_2$CuO$_4$-based superconductors. It is evident that $T_m$ will be a
maximum locally around $x \simeq \epsilon \simeq 1/8$, since for small
$\alpha$, admixtures of $n=4$ ($x < 1/8$) or $n=2$ ($x> 1/8$) ladders will
decrease the magnetic correlation lengths and hence will lower $T_m$.
However, there is no obvious reason why other odd-width ladder arrays
would not also be favorable.  We note that for $x < 0.05 = 1/20$ the spin
correlations are commensurate \cite{Endoh92,Kastner98,Yamada98} so that
the uniform stripe model does not apply. Thus, the other relevant
odd-width ladder arrays are $n=7$, corresponding to $x=\epsilon=0.0625$,
and $n=5$, corresponding to $x=\epsilon=0.083$. We have carried out some
limited calculations for $n=5$ and, as one would intuit from the results
of Greven {\it et al.} \cite{Greven96} for isolated ladders, at a given
temperature the correlation lengths for arrays of $n=5$ ladders are
somewhat larger than those for $n=3$ arrays. For perfectly ordered charge
stripes, $T_m$ should be correspondingly higher. In fact, the opposite
seems to be true, although more experimental data are required to document
this completely. At the minimum, from the inelastic neutron scattering
studies of Yamada {\it et al.}, \cite{Yamada98} we know that the
low-energy dynamical coherence length at low temperatures for $x \simeq
\epsilon \simeq 1/12$ is shorter than that for $x \simeq \epsilon \simeq
1/8$. 

Clearly, therefore, if this stripe model is to describe the real
La$_2$CuO$_4$-based superconductors, then some further physics is
required. At least three different factors could act to reduce the charge
and concomitantly the spin correlation lengths at lower doping. First, as
evidenced by the commensurate spin structure for $x \alt 0.05$, pinning of
the stripes becomes much more important at lower dopings.  Second, as the
charge stripe separation grows, the stripe-stripe interaction energy will
decrease and, correspondingly, stripe positional fluctuations will grow.
These would in turn inhibit the development of the spin correlations. 
Third, as discussed above, it is believed that at $x=0.05$ the stripes
switch from being approximately along the Cu--O--Cu--O direction to being
along the diagonal direction, \cite{Wakimoto98} that is from $(1 \; 0)$ to
$(1 \; 1)$. Presumably, as the doping decreases towards 0.05, fluctuations
into the diagonal stripe phase could occur and thus would shorten the spin
correlation length. These ideas are, of course, purely heuristic. Some
real theoretical calculations plus further experiments are clearly
required to put the model on a firmer basis. 

We have assumed that the charge stripes are site-centered rather than
bond-centered.  \cite{White98a} The basic reason for this is that, at
least within our simplified model, we are unable to match at all the
observed experimental trends with a bond-centered model. This may be
easily seen as follows. First, if the spins across the charged stripe
bonds are coupled antiferromagnetically, then the basic commensurate
antiferromagnetic situation is unaltered. Second, if, as originally
argued by Aharony {\it et al.}, \cite{Aharony88} holes on the oxygens
induce a strong ferromagnetic coupling between the two neighboring copper
spins, then one has effective $S=1$ chains. Thus, for the specific
case of $\epsilon=1/8$, the system breaks up into alternating spin-1
chains and two-leg ladders. Both the spin-1 chain ($\Delta \approx 0.4J$)
and the two-leg ladder ($\Delta \approx 0.5J$) have large spin gaps and
short correlation lengths, so that this system only orders at
extremely low temperatures, if at all. On the other hand, for $x=1/10$ or
$x=1/6$, one has single chains or three-leg ladders respectively in
between the spin-1 chains. Hence the correlation length along the chain
diverges at low temperature, making a spin density wave ordering
possible.  The difficulty for a bond-centered model then is that one
has facile ordering at $x=1/10$ and $x=1/6$, but not at $1/8$, which is
the opposite of the general trend observed experimentally. Therefore, we
are required within the context of our model to adopt site-centered
stripes in order to obtain sensible behavior around $x=1/8$ doping regime.
It would clearly be interesting to see if this same difficulty occurs for
the $t-J$ model as discussed by White and Scalapino.\cite{White98a}

Manifestly, more experiments are needed to address many unresolved
features of the stripe model. First, one needs to understand in detail the
doping dependence of the spin density wave ordering in the
La$_{2-x}$Sr$_{x}$CuO$_4$ and related systems. Second, better
characterization of the charge stripes themselves, including both the
incommensurabilities and the correlation lengths is crucial.  Third,
unambiguous determination of the charge ordering temperatures is
essential, since it is still not certain as to whether charge ordering
drives the spin ordering, or vice versa. \cite{Zachar98} Since neutrons
scatter from the small nuclear displacements induced by the modulated
charge density, the charge ordering peaks observed via neutron scattering
are extremely weak. Electron diffraction is, of course, especially
sensitive to the charge modulation, but it is only useful for surfaces or
very thin films which may differ from their bulk counterparts. X-ray
scattering seems to be a logical choice to study the charge modulation in
these materials. In their study of
La$_{1.48}$Nd$_{0.4}$Sr$_{0.12}$CuO$_4$, Zimmermann {\it et al.}
\cite{Zimmermann98} used high-energy x-ray (100 keV) to investigate charge
stripes, confirming results from neutron scattering experiments.
\cite{Tranquada} They were able to determine peak widths at various
temperatures with somewhat better precision than that of neutron
scattering; however, a much higher resolution x-ray study is necessary to
compare quantitatively the correlation length data with various
theoretical predictions. Generally, it is important to determine the
nature of the charge-charge stripe correlations in both the normal and
superconducting states.

In summary, we have studied the magnetism arising from the charge and spin
stripe order in monolayer cuprate superconductors. Quantum Monte Carlo
simulations have been carried out on a simplified model, which is
essentially weakly coupled insulating spin ladders. Our calculations are
consistent with recent results of spin density wave ordering and dynamical
spin fluctuations in La$_{2-x}$Sr$_{x}$CuO$_4$ and La$_{2}$CuO$_{4+y}$ for
doping and incommensurabilities near 1/8.  We show that the periodicity of
the incommensurate spin order is not necessarily twice that of the charge
order. The behavior at lower doping remains problematic. We have also
studied in detail the quantum critical behavior in coupled four-leg
ladders.

\acknowledgements{We would like to thank V. J. Emery, M. Greven, T. Imai,
S. A.  Kivelson, and J. M. Tranquada for invaluable discussions. We thank
S. A. Kivelson especially for detailed comments on this manuscript. The
present work was supported by the US-Japan Cooperative Research Program on
Neutron Scattering. The work at Tohoku has been supported by a
Grant-in-Aid for Scientific Research of Monbusho and the Core Research for
Evolutional Science and Technology (CREST) Project sponsored by the Japan
Science and Technology Corporation. The work at MIT was supported by the
NSF under Grant No. DMR97-04532 and by the MRSEC Program of the National
Science Foundation under Award No. DMR98-08941. The work at Brookhaven
National Laboratory was carried out under Contract No. DE-AC02-98CH10886,
Division of Materials Science, U. S. Department of Energy.
}

\end{document}